\documentclass[12pt]{iopart}

\usepackage{hyperref}
\newcommand{\be}{\begin{equation}}
\newcommand{\ee}{\end{equation}}
\newcommand{\bea}{\begin{eqnarray}}
\newcommand{\eea}{\end{eqnarray}}

\usepackage{graphicx}

\begin{document}

\title[Interpretations of cosmic expansion: anchoring conceptions and misconceptions]{Interpretations of cosmic expansion: anchoring conceptions and misconceptions}

\author{M P\"ossel$^1$}

\address{$^1$Haus der Astronomie and Max Planck Institute for Astronomy, K\"onigstuhl 17, 69117 Heidelberg, Germany}

\ead{poessel@hda-hd.de}

\begin{abstract}
Teaching cosmology at the undergraduate or high school level requires simplifications and analogies, and inevitably brings the teacher into contact with at least one of the pedagogical interpretations of the expanding universe. The by far most popular interpretation holds that galaxies in an expanding universe are stationary, while space itself expands and thus causes the growing distances that characterize cosmic expansion. The alternative relativistic explosion interpretation regards cosmic expansion as a pattern of (relativistic) galaxy motion. The aim of this article is to discuss the two competing interpretations from the perspective of potential student preconceptions, taking into account both beneficial anchoring conceptions and potentially harmful preconceptions that can lead to misconceptions. 
\end{abstract}

\vspace{2pc}

\noindent{\footnotesize This is the version of the article before peer review or editing, as submitted by the author to the journal {\em Physics Education}. IOP Publishing Ltd is not responsible for any errors or omissions in this version of the manuscript or any version derived from it. The Version of Record is available online at \href{https://doi.org/10.1088/1361-6552/aba3b1}{https://doi.org/10.1088/1361-6552/aba3b1}}
%
%
%

\section{Introduction}

At the graduate level, teaching about the expanding universe can rely on the appropriate mathematical tools of general relativity. There, the central mathematical object is the metric describing the geometry of Friedmann-Lema\^{\i}tre-Robertson-Walker (FLRW) spacetimes, from which all the basic properties of expanding universes can be derived. Together with the Friedmann equations, which state how the energy and matter content of the universe determines the evolution of cosmic expansion, they provide the foundation for our modern cosmological models. 

Where these mathematical tools are not available, notably at the undergraduate or high school level, teaching cosmology must rely on simplified models and analogies --- something that is true more generally when teaching about general relativity  \cite{kerstingGravityWarpedTime2020,posselRelativelyComplicatedUsing2018}. Unavoidably, such simplifications come with conceptual baggage. In cosmology, they must necessarily rely on an {\em interpretation} of cosmic expansion, that is, on a conceptual framework that provides elementary descriptions such as ``galaxies are moving away from each other.''

The most common such framework is the {\em expanding space interpretation}, which posits that galaxies participating in cosmic expansion are at rest, but that space between them is expanding. An alternative is the {\em relativistic motion interpretation}, which regards cosmic expansion as a pattern of (relativistic) galaxy motion in space. There is a considerable body of literature on the merits and problems of each interpretation \cite[and references therein]{narlikarSpectralShiftsGeneral1994a,davisExpandingConfusionCommon2004,francisExpandingSpaceRoot2007,liebscherSpacetimeCurvatureRecession2007,bunnKinematicOriginCosmological2009,posselCosmicEventHorizons2019}, but the discussion has mostly centered on the physical and mathematical aspects. The aim of this article is to examine the two interpretations from a complementary perspective, focussing on the connections with basic preconceptions that pupils are likely to bring to cosmology from everyday experience as well as from high-school and university-level physics. 

Such connections are of interest for teaching since preconceived knowledge can lead to misconceptions, but it can also aid understanding through ``anchoring conceptions'' in the sense of Clement, Brown \& Zietsman: elements of ``an intuitive knowledge structure that is in rough agreement with accepted physical theory,'' which allow for anchoring new material in learners' existing frameworks of knowledge \cite{clementNotAllPreconceptions1989}. While the present work does not include an empirical component, the results suggest ways of extending existing studies of student's cosmological conceptions and misconceptions \cite{wallaceStudyGeneralEducation2011,wallaceStudyGeneralEducation2012c,aretzDevelopmentEvaluationConstruct2017} by explicitly taking into account the different roles such preconceptions play for the two competing interpretations.

\section{Interpretations of cosmic expansion}

Modern cosmological models describe a family of idealised galaxies, said to be ``in the Hubble flow,'' in a universe that is taken to be  homogeneous on average. At each point in space, the Hubble flow defines a local standard of rest.
Pick out two arbitrary Hubble-flow galaxies, and their distances will change over time in proportion to a universal cosmic scale factor, often called $a(t)$. Although there are re-collapsing FLRW spacetimes, we will in the following concentrate on cosmic expansion, where $a(t)$ grows over time. An immediate consequence is the Hubble-Lema\^{\i}tre relation:  At the present time, the {\em recession velocity} $v_{rec}$, that is, the change over time of the distance $d$ between two Hubble-flow galaxies, is related to $d$ as 
\be
v_{rec}= H_0\cdot d,
\label{HubbleLemaitreRelation}
\ee
with $H_0$ the Hubble constant. Edwin Hubble's empirical version, published in 1929, was instrumental for the acceptance of the concept of an expanding universe in cosmology. With the FLRW spacetimes formulated by Friedmann and Lema\^{\i}tre, and refined later by Robertson, Walker and others, it became clear that for our own universe, general relativity predicts a time $t_{ini}$ in the past where $a(t_{ini})=0$: the big bang singularity. Physical descriptions of the big bang phase following the era directly after $t_{ini}$, where the universe was filled with a hot and dense plasma, make up a major portion of modern cosmological research.

The most common framework for interpreting FLRW spacetimes is the {\em expanding space interpretation}. An excellent review can be found in the seminal papers of Davis and Lineweaver, which are centered around popular misconceptions of cosmic expansion \cite{davisExpandingConfusionCommon2004,lineweaverMisconceptionsBigBang2005}. A thorough treatment can also be found in the book by Harrison \cite{harrisonCosmologyScienceUniverse2000}. At the core of this interpretation is the notion that galaxies are at rest in space. Changing inter-galaxy distances, then, are not due to galaxy motion {\em through} space. Instead, they are the consequence of space between the galaxies expanding. 

In the expanding space interpretation, light from distant galaxies is redshifted ``[b]ecause expanding space stretches all light waves as they propagate'' \cite{lineweaverMisconceptionsBigBang2005}. In fact, in FLRW spacetimes, the wavelength $\lambda_e$ with which the light is emitted by a distant galaxy at time $t_e$ and the wavelength $\lambda_r$ with which the light is received at our own galaxy at a later time $t_r$ are related to the cosmic scale factor as
\be
1+z\equiv \frac{\lambda_r}{\lambda_e} = \frac{a(t_r)}{a(t_e)},
\label{CosmicRedshiftScaleFactor}
\ee
where the left-hand equation defines the redshift $z$. This direct proportionality lends plausibility to the light-stretching interpretation.

Alternatively, in the {\em relativistic explosion interpretation}, galaxies {\em are} moving through space, and the cosmological redshift can be interpreted as a Doppler shift: as a consequence of observers in relative motion measuring the wavelength of the same light signal, and coming to different conclusions. Crucially, the relative radial velocity $v_{rad}$ of a galaxy moving directly away from us in the course of cosmic expansion is not the same as the recession velocity $v_{rec}$ defined by the Hubble-Lema\^{\i}tre law (\ref{HubbleLemaitreRelation}). It is a relativistic radial velocity derived using the general-relativistic notion of parallel transport \cite{narlikarSpectralShiftsGeneral1994a,posselCosmicEventHorizons2019}. In terms of this relativistic radial velocity, which is always subluminal, $v_{rad}<c$,
the cosmological redshift can be written using the same formula as the special-relativistic redshift for radial velocity,
\be
\frac{\lambda_r}{\lambda_e} = \sqrt{\frac{c+v}{c-v}},
\label{CosmicRedshiftDoppler}
\ee
which for FLRW spacetimes turns out to be an alternative way of writing the redshift formula (\ref{CosmicRedshiftScaleFactor}). 

The earliest reference to this interpretation appears to be due to Lanczos in 1923, based on calculations using de Sitter's cosmological solution \cite{lanczosUeberRotverschiebungSitterschen1923}, and thus predating the description of the more general FLRW spacetimes. Synge derives the Doppler redshift formula in his 1960 general relativity book \cite{syngeRelativityGeneralTheory1960}, and the result has been brought to the attention of more recent readers e.g.\ by Narlikar \cite{narlikarSpectralShiftsGeneral1994a} and by Bunn and Hogg \cite{bunnKinematicOriginCosmological2009}. 
A sign of the relative obscurity of this interpretation is that variations thereof have been re-discovered, independently, by different authors over the past decades \cite{bolosLightlikeSimultaneityComoving2006a,liebscherSpacetimeCurvatureRecession2007,kayaHubbleLawFaster2011,rebhanCosmicInflationBig2012a}.

It should be stressed that the disagreement between the two interpretations does not extend to the underlying mathematics of FLRW spacetimes, nor to the basic interpretation of Hubble-flow world-lines. Ask a question about measurable quantities, and both interpretations must give you the same answer. In consequence, the debate between the competing interpretations is at its core one of pedagogy: Which of the interpretations are more helpful for learners' understanding of cosmic expansion? That question can be rephrased in terms of learners' possible preconceptions: Which interpretation benefits most from concepts that learners already know? Which preconceptions can serve as anchoring conceptions? As soon as we describe cosmic expansion beyond abstract statements about increasing distances and wavelengths, using terms such as galaxies ``moving,'' or ``space getting stretched,'' those terms carry associations of their own, carried over from their meanings in everyday conversation, and also their role in other areas of physics. Which of these associations help, and which are potentially confusing?
 
\section{Moving vs. getting carried away}

\begin{figure}[htbp]
\begin{center}
\includegraphics[width=0.5\textwidth]{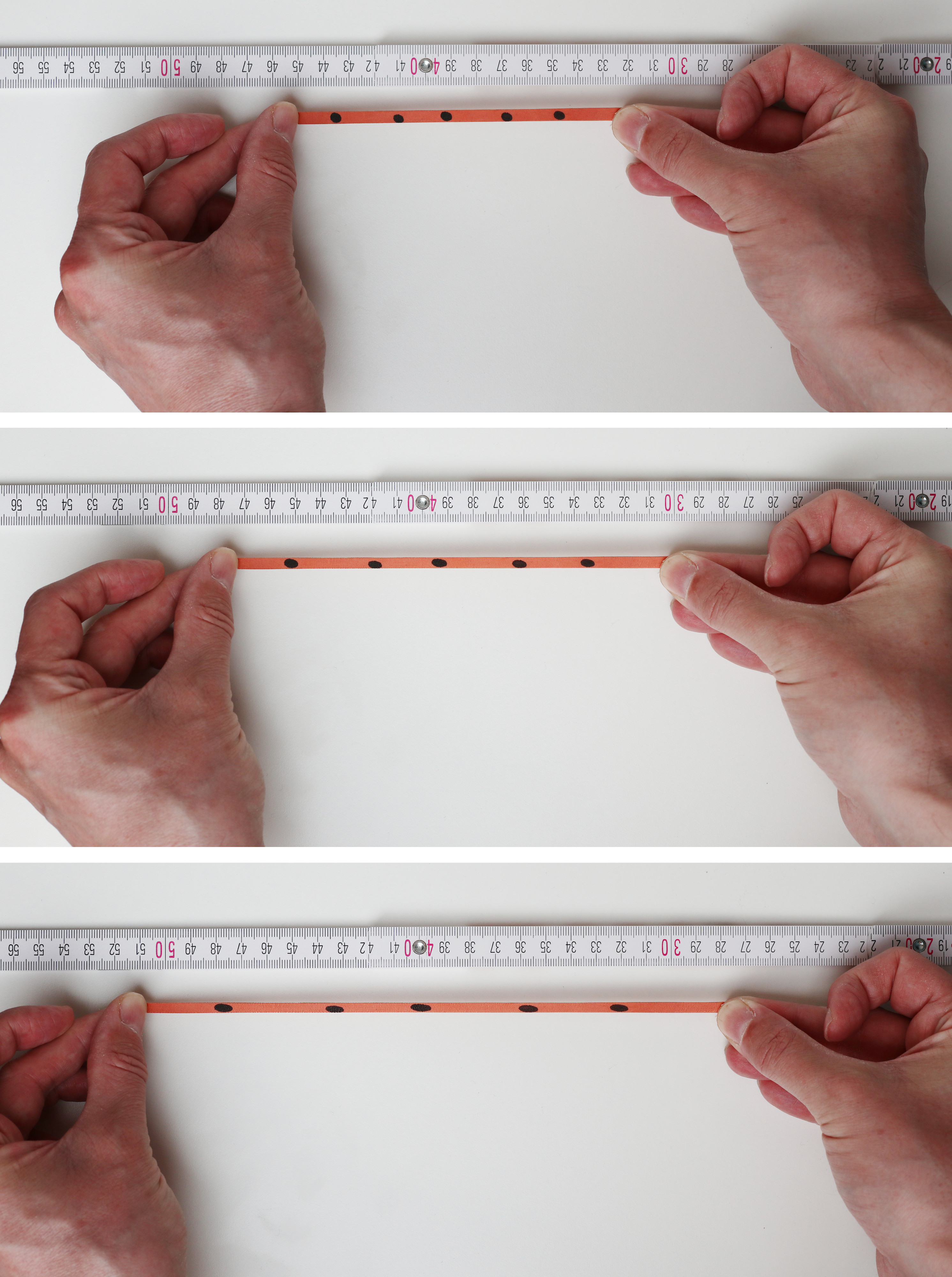}
\caption{One-dimensional rubber band universe getting stretched over time (top to bottom). Galaxies are represented by dots}
\label{CosmosModels1D}
\end{center}
\end{figure}

\begin{figure}[htbp]
\begin{center}
\includegraphics[width=\textwidth]{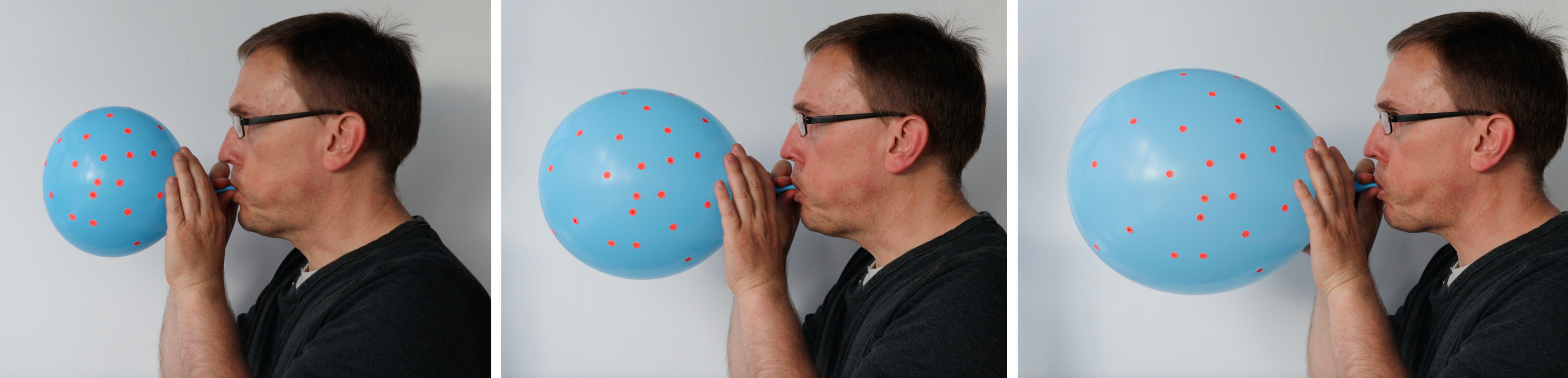}
\caption{A balloon, whose surface represents a two-dimensional universe, being inflated (left to right). Galaxies are represented by stickers}
\label{CosmosModels2D}
\end{center}
\end{figure}

A considerable advantage of the expanding space interpretation is its close relation with what I will refer to collectively as {\em expanding-substrate models}: teaching models in which the space of an expanding universe is represented by a medium or substrate. 
These include a stretching rubber band which represents a one-dimensional universe, with painted-on dots for galaxies, Fig.~\ref{CosmosModels1D}, 
and an inflating rubber balloon with stickers for galaxies, as a two-dimensional cosmos, Fig.~\ref{CosmosModels2D}. The models allow for (limited) quantitative measurements of the Hubble-Lema\^{\i}tre relation \cite{priceCosmologicalExpansionClassroom2001,fraknoiUniverseYourFingertips2011}.
We ourselves have no everyday experience of space expanding, but there are relevant preconceptions directly related to these expanding-substrate models. Why, for instance, do the distances between galaxy-stickers on the rubber balloon change? They are being (a) pulled along by the expanding surface, which requires (b) that the surface itself is expanding and (c) that the stickers are affixed to the surface. 

How are these preconceptions-by-proxy related to the underlying physics of cosmic expansion? Part of the answer, culminating in the question of to what extent it even makes sense to ascribe properties to space directly, would lead us into rather deep philosophical waters, related to Einstein's version of Mach's principle \cite{huggettAbsoluteRelationalTheories}, far beyond the scope of this article. Instead, consider a much simpler approach: Real galaxies can have non-zero {\em peculiar velocities} relative to their local Hubble-flow standard of rest. For example, when the center of mass of a galaxy cluster follows the Hubble flow, individual galaxies orbiting within the cluster are in motion relative to that center. At any point in space, a galaxy with peculiar motion up to the local speed of light is just as compatible with cosmic expansion as a galaxy in the Hubble flow. Whatever the ``medium'' that expands in this interpretation, it cannot be one to which galaxies can be said to be ``affixed.''

What about galaxies getting ``pulled along''? Consider a galaxy which has just the right peculiar velocity to be momentarily at rest relative to our own galaxy. If you expect that galaxy to begin moving away from us, getting pulled along with the Hubble flow, you would be wrong. Depending on the matter content of the model universe, and on cosmic time, even in an expanding universe, that galaxy might instead get pulled {\em towards} us \cite[and references therein]{barnesJoiningHubbleFlow2006a}. What happens depends only on the second derivative of the scale factor, not on the first derivative, in other words: What happens is independent of how fast our cosmos is expanding at this particular moment. 

In the relativistic-explosion interpretation, on the other hand, several preconceptions concerning motion can serve as helpful anchoring conceptions for these same situations. On the simplest level, relative motion is what we call it when distances between objects increase, and we routinely determine an object's state of motion by how its distances to salient points in its immediate environment are changing. On a more advanced conceptual level, we know from classical mechanics that dynamical interactions are a matter of second time derivatives, of accelerations, while first derivatives and positions can be chosen freely at any fixed moment in time, serving as initial values for the differential equations which govern the dynamics of a system. Once acquired, this knowledge can be used as an anchoring conception for understanding why galaxies can have non-zero peculiar velocities, and also to understand the behaviour of a galaxy outside the Hubble flow: whether such a galaxy will begin to accelerate towards us or away from us is a question of dynamics, depending on whether the gravitational interactions in the expanding universe are predominantly attractive (ordinary and dark matter, radiation) or repulsive (dark energy). But dynamics does not restrict the momentary velocity that a galaxy can have --- when setting up a particular model situation, that velocity can be chosen freely, as an initial condition. Initial conditions are separate from dynamics; peculiar velocities are not restricted by cosmic expansion.

In addition, peculiar velocities and the relativistic radial velocities due to cosmic expansion follow special-relativistic preconceptions about how velocities add up, namely by the special-relativistic addition formula \cite{emtsovaVelocitiesDistantObjects2019,posselCosmicEventHorizons2019}.

The differences between the two interpretations are particularly interesting for bound systems. If space itself is expanding, then are atoms, planetary systems or galaxies expanding as well? Naively, if all of space is getting bigger, that should also hold for the space between, say, sun and earth. The relevant calculations show that, indeed, bound systems react to cosmic expansion by shifting their equilibrium sizes (commonly by undetectable amounts), but again, the result only depends on the acceleration or deceleration of cosmic expansion \cite{cooperstockInfluenceCosmologicalExpansion1998,giuliniDoesCosmologicalExpansion2014}. For what happens to bound systems, it is irrelevant how quickly space expands at any given instant in time. This is hard to reconcile with any interpretation that attempts to understand the situation as an equilibrium between the expansion of space and binding forces. After all, the most immediate manifestation of the expansion of space, namely how quickly distances are increasing right now, plays no role whatsoever. 

In the relativistic explosion interpretation, the notion of initial conditions once more provides an anchoring conception: only the acceleration component matters, because only accelerations are important for dynamical considerations, including equilibrium states. Instantaneous speeds only enter as initial conditions. Systems remain bound if their initial conditions, which determine the system's total energy, are suitable for a bound state, and if there is an equilibrium between the binding force (e.g. gravitational or electrostatic) and the gravitational acceleration associated with cosmic expansion. 

\section{Light propagation and the cosmological redshift}

\begin{figure}[htbp]
\begin{center}
\includegraphics[width=0.8\textwidth]{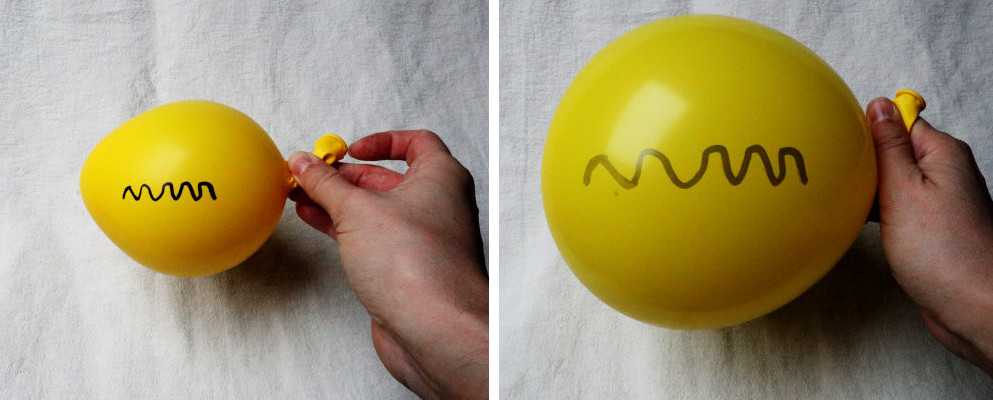}
\caption{When a balloon, whose surface represents a two-dimensional universe, is inflated (left to right), light-waves drawn on the balloon surface are stretched}
\label{WaveStretching}
\end{center}
\end{figure}

The most prominent property of light in an expanding universe is the cosmological redshift (\ref{CosmicRedshiftScaleFactor}) for light reaching us from distant Hubble-flow galaxies. In the expanding space interpretation, that redshift can be visualized using expanding-substrate models. The wavelength of a wave drawn onto the substrate, as in Fig.~\ref{WaveStretching}, will stretch in the same way as distances between Hubble-flow galaxies. 
When it comes to the model-based preconceptions, this correct stretching behaviour is a great plus. The downside is that the stretching amounts to parts of the wave getting ``pulled along,'' which in turn depends on the wave structure being affixed to the substrate, which in turn is inconsistent with the light moving freely, and at great speed, relative to the substrate.

In the relativistic explosion interpretation, the cosmological redshift is a Doppler shift \cite{narlikarSpectralShiftsGeneral1994a,bunnKinematicOriginCosmological2009}, its magnitude given by the special-relativistic formula (\ref{CosmicRedshiftDoppler}). This establishes students' pre-existing knowledge of both the special-relativistic and the classical Doppler effect as anchoring conceptions. It also resolves another potential conflict: If photons from distant galaxies arrive at our own galaxy with less energy than when they were emitted, where does that energy go? At first glance, that would appear to be in direct contradiction to the conservation of energy.
But if a Doppler effect is the explanation, then there is no such concern, since the two energy values refer to different reference frames.

In the expanding space interpretation, that apparent energy loss is less readily explained. Those texts that address the problem explicitly link it to the more general impossibility of defining global energy conservation in general relativity \cite{harrisonCosmologyScienceUniverse2000,davisUniverseLeakingEnergy2010}. A drawback of this solution is that the problem is also present in the zero-density limiting case of an empty FLRW spacetime, the so-called Milne universe \cite[sec.~16.3]{rindlerRelativitySpecialGeneral2001}, where physics is governed by special relativity, and where there most certainly {\em is} global energy conservation. 

In both special and general relativity, light propagation defines an absolute cosmic speed limit in the sense that no material object or signal can overtake a light signal. This is where the distinction between the recession speed, defined as in (\ref{HubbleLemaitreRelation}), and the relativistic radial velocity that is central to the relativistic explosion interpretation is crucial. Recession speeds become superluminal for distant galaxies. This appears to contradict students' preconceptions from special relativity, of the speed of light as a cosmic speed limit, and the apparent contradiction has been cited as key motivation for the expanding space interpretation: The differentiation between cosmic expansion as due to ``expanding space'' on the one hand, and ``galaxy motion through space'' on the other, is meant to address this conflict \cite{davisExpandingConfusionCommon2004}.

Relativistic radial velocities in the relativistic explosion interpretation never exceed the speed of light. From this perspective, superluminal recession speeds in (\ref{HubbleLemaitreRelation}) are an artefact, caused by a particular coordinate choice: The cosmic time coordinate ties together local clock rates in Hubble-flow galaxies, but clocks in relative motion tick at different rates, as we know from special relativity. Combining them into an overarching time coordinate, and using that coordinate to determine one-way speeds, leads to unphysical results. Students who have been on longer international flights know a closely related phenomenon: If your flight leaves Amsterdam at 15:00 local time and arrives in New York at 17:00 local time, this does not amount to a flight time of 2 hours, and corresponding average ground speed of 3000 km per hour.

The relativistic explosion interpretation can also readily explain a certain types of cosmological horizon with reference to the simple realisation that a slower-moving object following a faster-moving object will fail to catch up. Applied to the relativistic radial velocity, this gives a plausible explanation for why light from some distant regions can never reach us. Any boundary between regions whose light can reach us and regions whose light cannot, is called a horizon. In some FLRW spacetimes, there is a type of cosmological horizon that can be defined as the boundary where the relativistic radial velocity of Hubble-flow galaxies relative to our own galaxy approaches the speed of light --- so light sent in our direction from those galaxies cannot catch up with us \cite{posselCosmicEventHorizons2019}. Explanations for the same kind of cosmological horizon in the expanding space interpretation, on the other hand, need to include an explanation of why this simple argument is {\em not} true for recession speeds \cite{neatIntuitiveApproachCosmic2019}.

Last but not least, preconceptions from special relativity can help students understand why a relativistic explosion need not contradict the homogeneity of our universe. This is easiest to see for the Milne universe, where each galaxy corresponds to an inertial observer, all on an equal footing because of the same relativistic effects that combine in special relativity to yield the constancy of the speed of light for all inertial observers \cite[sec.~16.3]{rindlerRelativitySpecialGeneral2001}.

\section{Conclusion}

Each of the two main interpretations of cosmic expansion is connected with potential student preconceptions --- from everyday notions about motion to more advanced concepts that one encounters in physics courses.

For the expanding space interpretation, most of the relevant preconceptions are associated with the expanding-substrate models. This is a considerable plus both for teaching about the basic pattern of expansion with a universal scale factor and for light waves ``getting stretched'' as space expands. Beyond that, the preconceptions are more likely to hinder than aid understanding, as we have seen regarding peculiar velocities, long-term behavior of galaxies outside the Hubble flow, and bound systems.

For the relativistic explosion interpretation, preconceptions about motion serve learners well in understanding the cosmological redshift, galaxies outside the Hubble flow, bound systems, and a specific kind of cosmological horizon. Some of the preconceptions involved are very basic (motion changes distances), while others are more advanced, such as the Doppler effect, or the distinction between dynamics and initial conditions. Two disadvantages are that the interpretation is less closely related to the expanding-substrate models, and that any rigorous derivation of the relativistic radial speed is far beyond the level of high school or undergraduate cosmology teaching.

In teaching about cosmology, we should take these preconceptions into account. With the systematic overview provided by this text, I hope to raise awareness among astronomy educators of the potential benefits, but also the potential pitfalls involved.

\section*{Acknowledgements}

I am grateful to Markus Nielbock and Anna P\"ossel for their help in creating the illustrations, and to Thomas M\"uller for his critical comments on an earlier version of this text.

\section*{References}

\bibliography{ExpansionConcepts}{}

\end{document}